\documentclass[osajnl,twocolumn,showpacs,superscriptaddress,10pt]{revtex4-1}
\usepackage{amsmath,amssymb,graphicx}

\begin{document}

\bibliographystyle{unsrt}

\title{Saturation of atomic transitions using sub-wavelength diameter tapered optical fibers in rubidium vapor}
\author{D.E. Jones}
\author{J.D. Franson}
\author{T.B. Pittman}
\affiliation{Physics Department, University of Maryland Baltimore
County, Baltimore, MD 21250}

\begin{abstract}
We experimentally investigate ultralow-power saturation of the rubidium D$_{2}$ transitions using a tapered optical fiber (TOF) suspended in a warm Rb vapor. A direct comparison of power-dependent absorption measurements for the TOF system with those obtained in a standard free-space vapor cell system highlights the differences in saturation behavior for the two systems. The effects of hyperfine pumping in the TOF system are found to be minimized due to the short atomic transit times through the highly confined evanescent optical mode guided by the TOF. The TOF system data are well-fit by a relatively simple empirical absorption model that indicates nanoWatt-level saturation powers.
\end{abstract}

\ocis{(190.4360) Nonlinear optics, devices; (300.6210) Spectroscopy, atomic; (300.6460) Spectroscopy, saturation; (350.4238) Nanophotonics.}

\maketitle


\section{Introduction}
\label{sec:intro}

The use of sub-wavelength diameter tapered optical fibers (TOFs) suspended in atomic vapors has recently been identified as a promising platform for ultralow-power nonlinear optics \cite{spillane08,pittman13}. 
TOFs allow the propagation of highly confined optical modes over interaction lengths that can be much longer than those achieved by focusing a free-space beam into an atomic vapor cell \cite{tong04}. 
Consequently, TOFs surrounded by atomic vapors have enabled various forms of nonlinear spectroscopy to be performed at significantly reduced power levels \cite{hendrickson10,salit11,lai13}. 
The situation is closely related to hollow-core photonic bandgap fibers (PBGFs) filled with atomic vapors, where record-setting ultralow-power nonlinear phase shifts have recently been observed \cite{venkataraman13}. 

Much of this nonlinear optics work has involved the rubidium  $5S_{1/2} \rightarrow 5P_{3/2} \rightarrow 5D_{5/2}$ two-photon ladder transition at 780 nm and 776 nm \cite{spillane08,hendrickson10,salit11,lai13,venkataraman13}. 
The ability to saturate the Rb D$_{2}$ transition at 780 nm with ultralow-powers is a good indicator of the overall strength of the nonlinearity in the system \cite{spillane08}. 
Roughly speaking, it provides a measure of the overlap of the confined optical mode with the atomic absorption cross-section \cite{demtroderbook}.  
As the incident power is increased, the ability of the system to absorb the radiation decreases because the excited state population saturates. 
Remarkably low-power ($\sim \,$nW level) saturated absorption has been observed using TOF's in Rb vapor \cite{spillane08,hendrickson10,lai13}, but the effect has not yet been studied in detail. 
Here we report detailed measurements comparing saturation effects for a TOF in Rb system with those observed using a free-space beam focused into a standard Rb vapor cell. 

The direct comparison of these two different systems highlights two main results: (1) modeling the saturation behavior in the TOF system is complicated by the unique waveguiding geometry, and (2) the effects of hyperfine pumping in the TOF system are negligible compared to those present in typical free-space vapor cell systems. 
A tangible outcome of this study is the establishment of a fairly simplistic empirical model that provides a reasonable way to estimate the saturation power for TOF in Rb systems.

Deriving accurate theoretical models of saturation effects in atomic vapors is notoriously difficult, even in the simplest setting of a free-space beam focused into a Rb vapor cell \cite{siddons08}. 
The beam profile is a critical parameter, and non-uniformities lead to spatial variations in the excited-state populations that typically require numerical methods \cite{stace10}. 
In addition, the Rb D$_{2}$ transition can not be approximated as a closed two-level system, and hyperfine pumping significantly affects the population dynamics \cite{smith04,sherlock09}. 
This is particularly important when the transit time of the thermal atoms passing through the optical beam is comparable to the hyperfine pumping time (typically $\sim \mu$s), in which case transient effects complicate the saturation behavior \cite{moon08a,moon08b,lindvall09,perrella12,sprague13}. 

The situation is far more complex for the TOF in Rb system, where the intensity profile of the guided evanescent mode interacting with the atoms is very non-uniform \cite{tong04}. 
In addition, collisions with the TOF itself may play a significant role  \cite{ghosh06,slepkov10}. 
The presence of the fiber also affects the velocity distribution of atoms interacting with the evanescent mode, and the very brief transit-time introduces a significant line-broadening mechanism \cite{spillane08,hendrickson10}. 
Consequently, the development of an accurate theoretical model of the saturation behavior is beyond the scope of this paper. 
Instead, we focus on a simple empirical model that best fits nonlinear absorption measurements over a full-range of input powers. 
The goal is to provide a reasonable way to quote estimates of the saturation powers for TOF in Rb systems.

\section{TOF in Rb system}
\label{sec:TOFinRb}

Figure 1 highlights the relevant features of the TOF waveguiding geometry. 
The blue area shows SEM measurements used to profile one of our optimized TOFs, and the red area shows the calculated electric field distribution in the waist region using the formalism of \cite{tong04}. 
When the TOF diameter is roughly half the wavelength of the propagating light, the TOF supports a guided $HE_{11}$ mode with a significant fraction of the field propagating as an evanescent wave outside the TOF itself (where it can interact with the surrounding Rb vapor). 
If the TOF diameter is made smaller, a larger fraction of the field is outside the fiber, but the cross-sectional mode area quickly blows up \cite{tong04}.

\begin{figure}[t]
\includegraphics[width=3.25in]{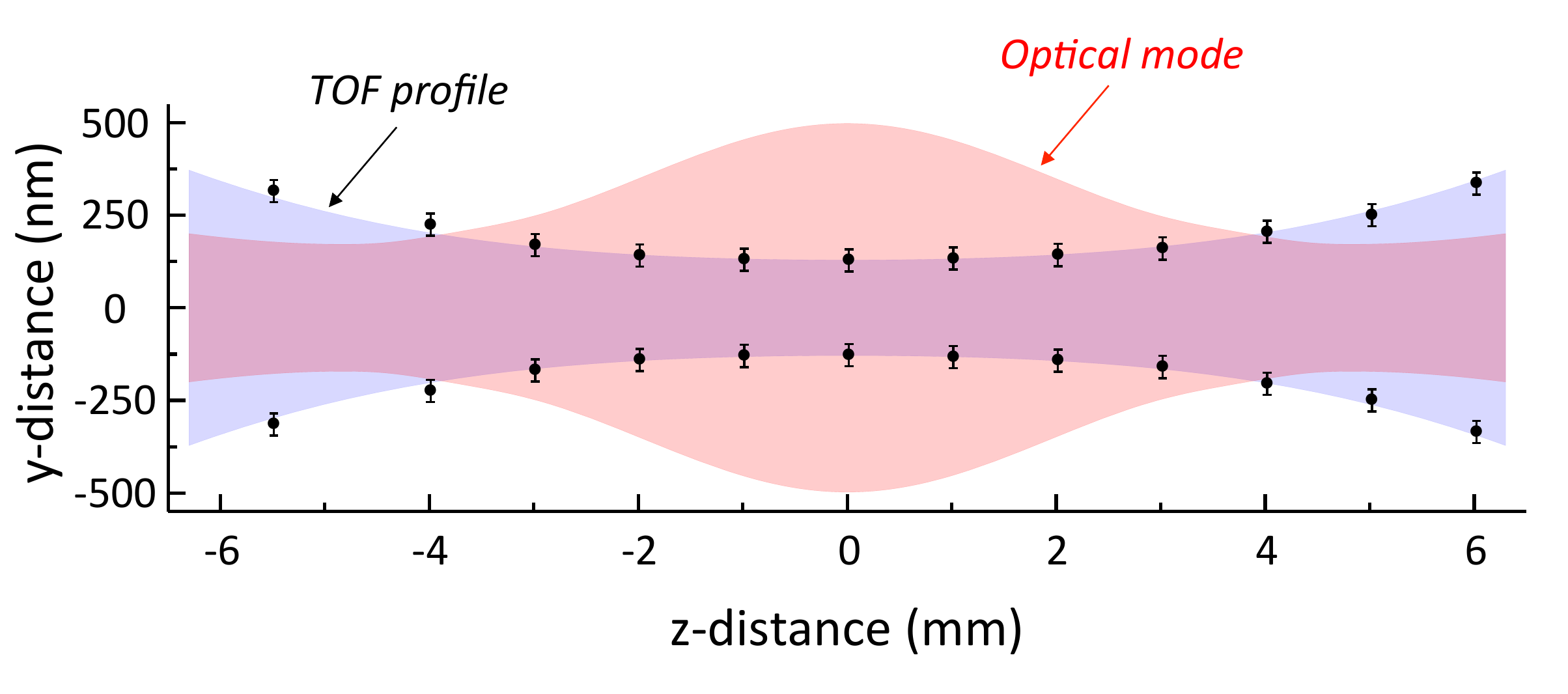}
\caption{(Color online) Overview of the TOF system. The data points are SEM measurements of the diameter used to profile the TOF (represented by the blue shaded area). The red shaded area represents a calculation of the $HE_{11}$ optical mode guided by the TOF (the mode is shown as the diameter in which half of the total power is confined) \protect\cite{tong04}.  Note that the aspect ratio of the figure is compressed by a factor of $10^{4}$; the key point is the propagation of a very tightly confined evanescent field over a very long interaction length in a surrounding Rb vapor.}
\label{fig:mode}
\end{figure}

Consequently, this trade-off leads to an optimal TOF diameter for nonlinear optics applications \cite{you08}.  
For the 780 nm light of interest here, the optimal diameter is about 320 nm, and the TOF shown in Figure 1 is near optimal. 
Note that the optical mode has a cross-sectional mode area on the order of $\sim\!1$ $\mu$m$^{2}$ propagating over a distance on the order of $\sim\!1$ cm. 
This confined geometry beats the Rayleigh limit for free-space focusing and enables the ultralow-power optical nonlinearities in this system.

It can be inferred from Figure 1 that atoms interacting with the strongest part of the optical field are also more likely to undergo collisions with the fiber itself. 
Wall collisions are known to be a significant homogeneous line-broadening mechanism in tightly confined waveguiding experiments \cite{ghosh06}.  
This is very different than the case of a typical free-space vapor cell experiment, where collisions with the cell walls are infrequent and occur far from the optical mode \cite{gorris97}.

\section{Experiment}

Figure 2 shows a simplifed block-diagram of the experimental setup. 
The TOF's were pulled using the flamebrush technique \cite{birks92} and installed in a vacuum system using a specialized TOF heating unit designed to minimize the accumulation of Rb on the TOF surface \cite{lai13}. 
The experiment was driven by a single fiber-coupled narrowband tunable diode laser at 780 nm. 
In order to make comparative measurements, the signal was split into two primary systems: the free-space vapor cell system and the TOF system. 
Independent in-line fiber variable attenuators (VA's) were used to accurately control the power being sent into each system over the wide dynamic range (nW to mW levels) needed for studying saturation behavior in the two systems. 

\begin{figure}[b]
\includegraphics[width=3.0in]{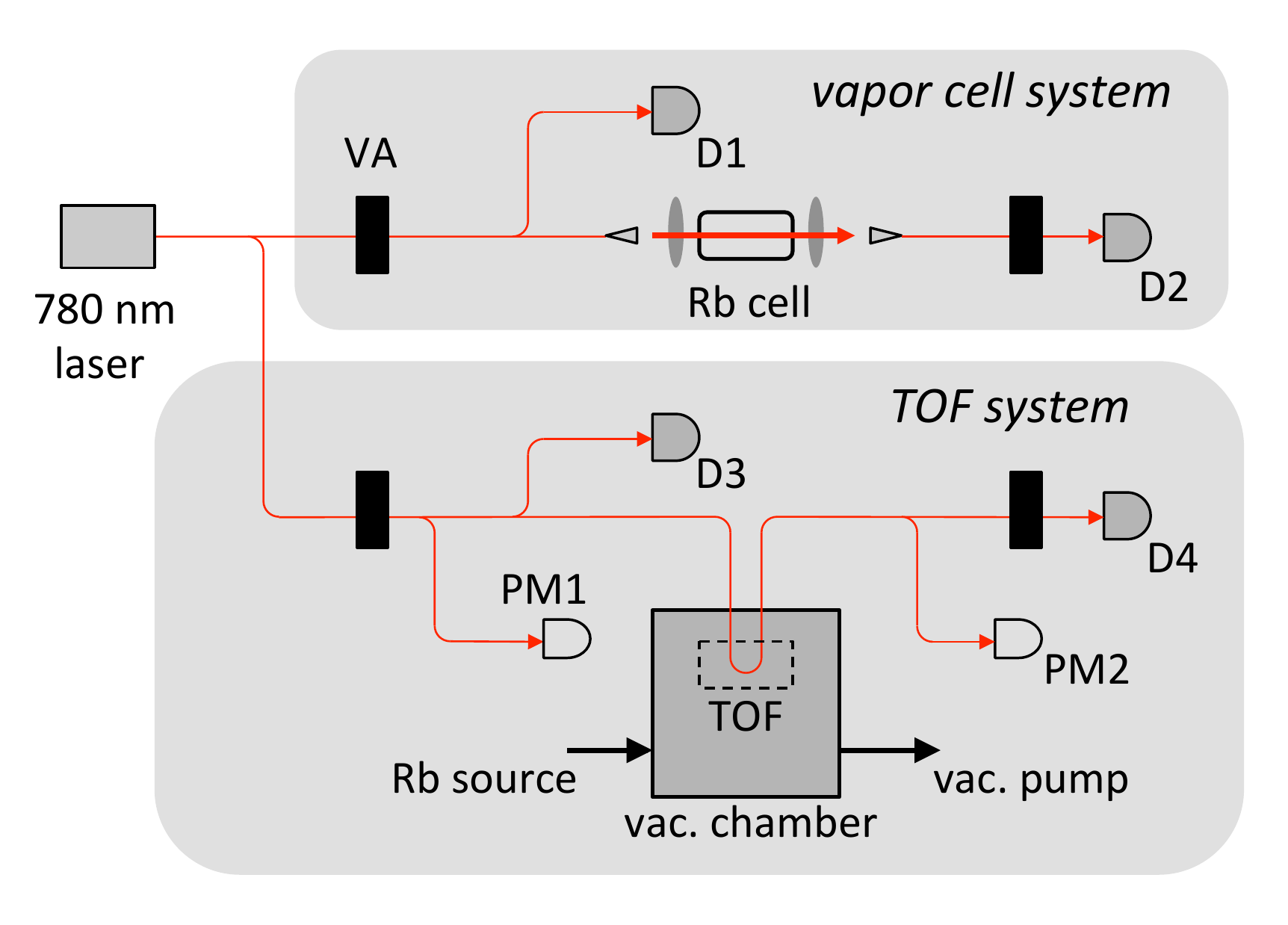}
\caption{(Color online) Block diagram of the experimental setup highlighting the vapor cell system and the TOF system. The entire apparatus is fiber-based (red lines), except for the free-space beam passing through the Rb vapor cell.  Detectors D1 (input) and D2 (output) are used to record transmission spectra for the vapor cell system, while detectors D3 and D4 are used for the TOF system. Variable attenuators (VA's) are used to control the input powers to the two systems. Additional details on the TOF vacuum system can be found in \protect\cite{lai13}.}
\label{fig:setup}
\end{figure}

The Rb density in the TOF vacuum system was controlled by heating a metallic Rb sample contained in a cold-finger attached to the main vacuum chamber (for additional details, see \cite{lai13}). 
Two auxiliary power meters (PM1 and PM2) were used to continuously monitor the overall transmission of the TOF system (typically $\sim 35\%$) during the experiments. 
The Rb density in the vapor cell was controlled by a simple oven. 
Lenses surrounding the Rb vapor cell were used to control the size of the free-space beam propagating through the cell.

Figure 3 shows typical Doppler-broadened absorption spectra obtained in the two systems when scanning the laser frequency over the D$_{2}$ line. 
In both cases, the powers are adjusted below the nominal saturation powers. 
The four primary dips (labeled 1-4) are due to the ground-state hyperfine splittings of the $^{85}$Rb and $^{87}$Rb isotopes illustrated in the level diagram of Figure 3(a). 

\begin{figure}[t]
\includegraphics[width=3.25in]{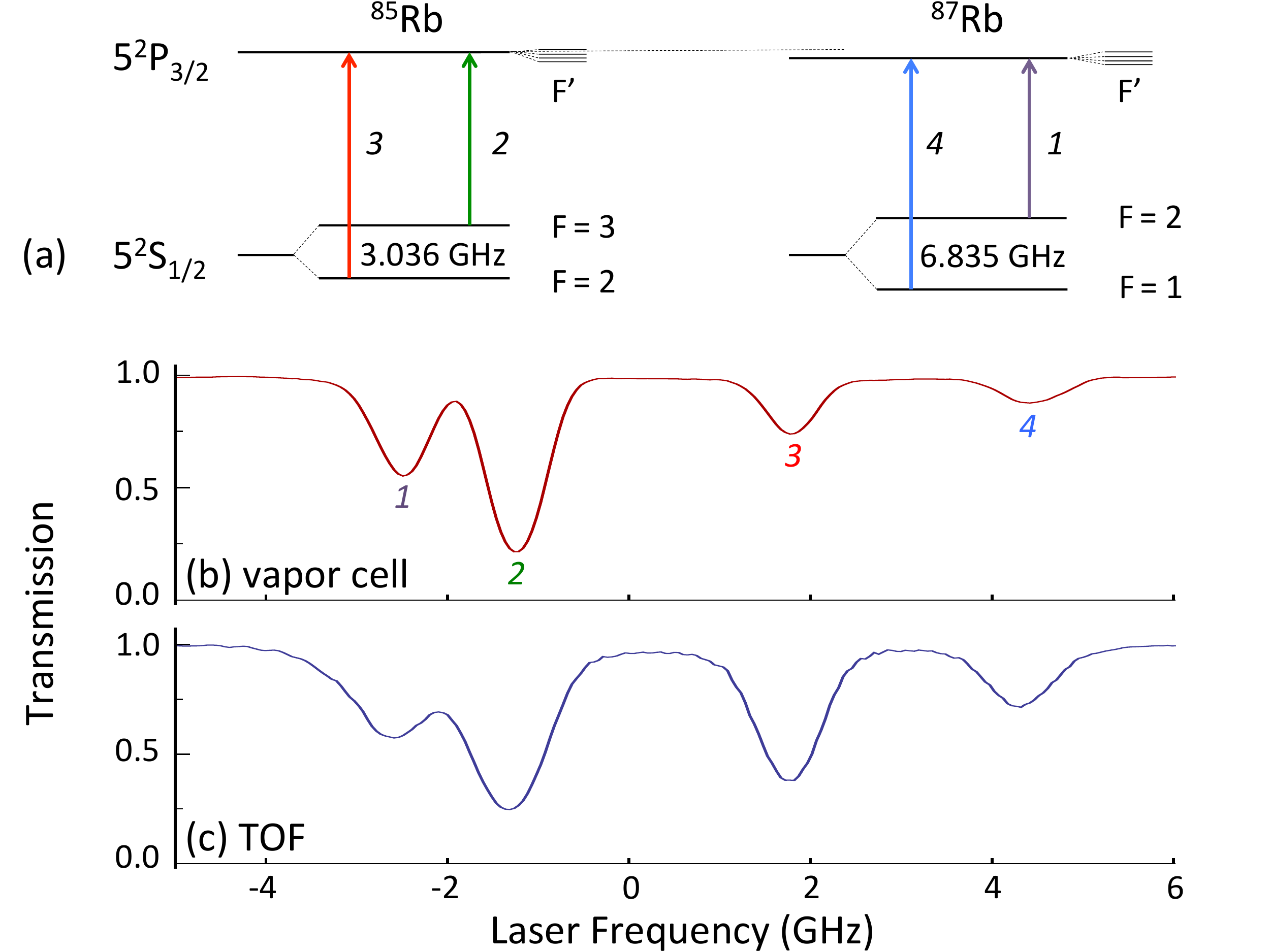}
\caption{(Color online) (a) Simplified energy level diagram for the relevant D$_{2}$ transitions in $^{85}$Rb and $^{87}$Rb; the excited state hyperfine splittings ($F'$) are closely spaced and not resolved due to Doppler broadening. For convenience, the transitions (and corresponding absorption dips) are labeled 1 - 4.   Plot (b) shows transmission spectra obtained in the vapor cell system, while plot (c) shows corresponding data simultaneously obtained in the TOF system. A comparison of (b) and (c) highlights the differences in transit time broadening and hyperfine pumping rates for these two systems.}
\label{fig:dips}
\end{figure}

Two key differences in the observed absorption spectra provide insight into the differences in saturation behavior of these two systems. 
First, the general lineshapes of the TOF dips are slightly different than those in the vapor cell (the dips are better fit by profiles with stronger Lorentzian contributions). 
This spectroscopic feature, which was first observed in \cite{spillane08}, is primarily a consequence of transit-time broadening due to the small optical mode area of the TOF system and collisions with the TOF itself \cite{watkins13}.
Second, and more importantly, the relative depths of dip 3 vs. dip 2 for $^{85}$Rb (and dip 4 vs. dip 1 for $^{87}$Rb) in each system are very different due to significantly reduced hyperfine pumping in the TOF system.


\section{Hyperfine pumping}
\label{sec:pumping}

Hyperfine pumping is known to significantly alter the saturation dynamics in Rb vapor cell experiments \cite{siddons08,stace10,smith04,sherlock09,moon08a,moon08b}. 
Figure 4 shows an example of the essential physics of this process and its relevance in measuring saturated absorption. 
Here, the laser is tuned to drive the $5S_{1/2}, F=2 \rightarrow 5P_{3/2}$ transition (labeled $|a\rangle \rightarrow |b\rangle$) in $^{85}$Rb (corresponding to dip 3), which experiences significant hyperfine pumping.  
If the excited state population $|b\rangle$  decays by spontaneous emission to the $F=3$ ground state $|c\rangle$, the population is effectively ``removed'' from the system and can no longer absorb the incident field. 

Figure 4 shows a plot of the time-scales of this effect for several field intensities. 
The dynamics were modeled using a standard rate equation approach for a single 3-level atom \cite{sherlock09}. 
The branching ratios for spontaneous emission from the excited state were calculated from the Clebsch-Gordan coefficients, assuming equal population of the magnetic sub-levels. 
For illustrative purposes, the initial population of the pumped state $|c\rangle$ was taken to be zero, and ensemble effects were neglected \cite{lurie12}. 

\begin{figure}[b]
\includegraphics[width=3.25in]{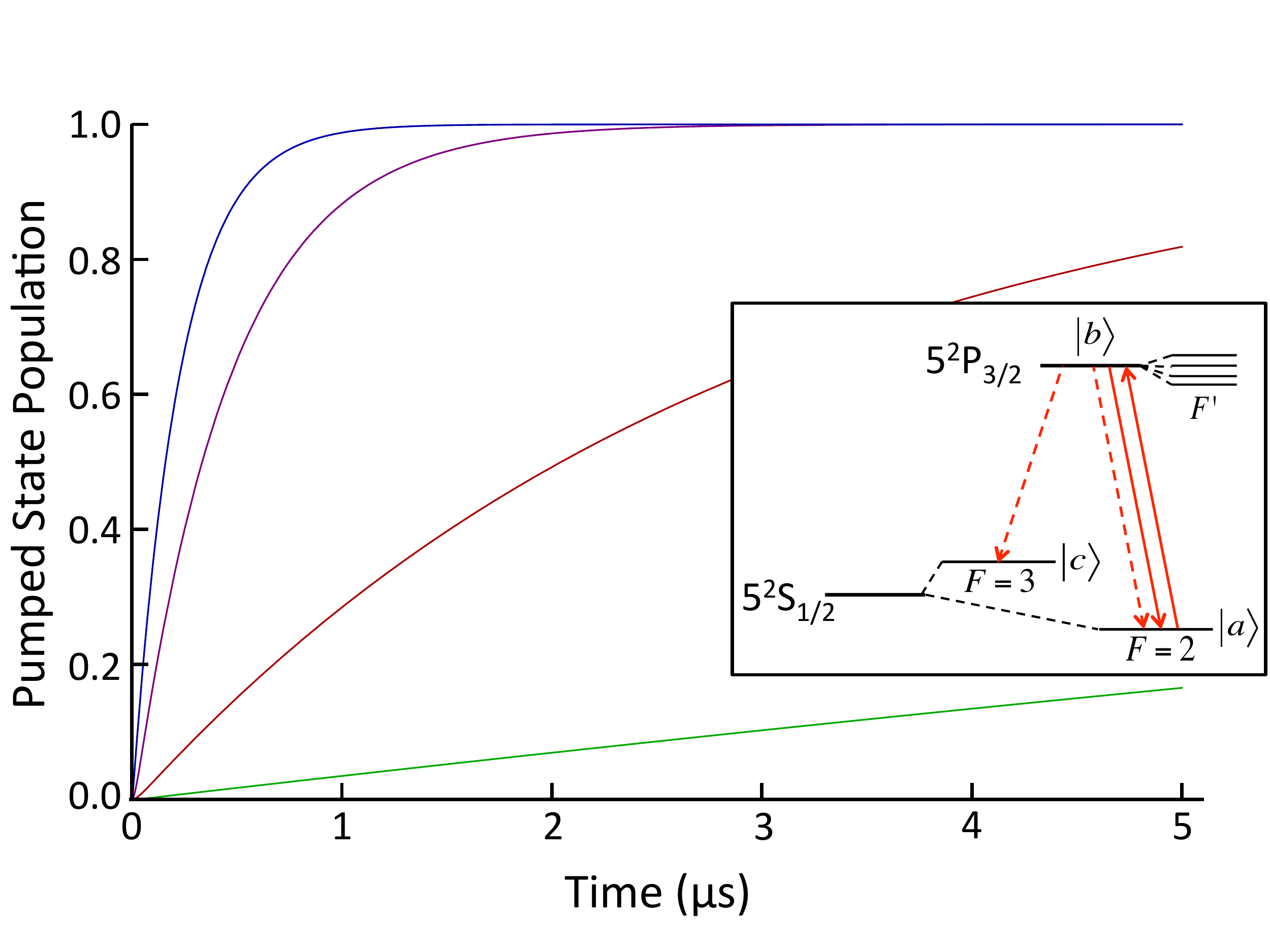}
\caption{(Color online) Example of calculated population vs. time for the hyperfine ground-state $|c\rangle$ while driving the $|a\rangle \rightarrow |b\rangle$ transition corresponding to dip 3 (see inset). The blue, purple, red, and green curves (left to right) correspond to driving intensities of 10, 1, 0.1, and 0.01 I$_{sat}$, respectively. Here I$_{sat}$ is given by the nominal Rb value of 1.6 mW/cm$^{2}$. In all cases, the key point is that the hyperfine pumping time is on the order of $\mu$s. This is comparable to the atomic transit times in vapor cell experiments, but much longer than the ns transit times in TOF experiments.}
\label{fig:population}
\end{figure}

The key point in Figure 4 is that the time-scale of Rb hyperfine pumping is on the order of a few $\mu$s. 
For typical room-temperature free-space vapor cell experiments using beam diameters on the order of $\sim$1 mm, this is comparable to the atomic transit time through the beam \cite{smith04}. 
Consequently, the effects of hyperfine pumping on apparent ``saturation'' of the transitions are significant \cite{moon08a,moon08b}. 
In contrast, the average transit time through the $\sim$1 $\mu$m size beam in the TOF system is on the order of a few ns, meaning the atoms have left the mode before significant hyperfine pumping occurs. 
For example, at intensities equal to $0.1I_{sat}$ and $I_{sat}$, only $\sim$.1\% and  $\sim$1\%, respectively, of the population has transferred from the initial state to the pumped state after 10 ns.

Figure 5 shows experimental measurements highlighting the effects of hyperfine pumping on apparent saturation for both the TOF and vapor cell systems. 
The data show the transmission level at each dip (labeled 1 - 4) as the power is increased towards the nominal saturation power. 
Summing over the dipole selection rules and oscillator strengths for the various hyperfine transitions indicates that dip 3 is more susceptible to hyperfine pumping than dip 2 for $^{85}$Rb , while dip 4 is more strongly affected than dip 1 for $^{87}$Rb \cite{siddons08}.

\begin{figure}[t]

\includegraphics[width=3.25in]{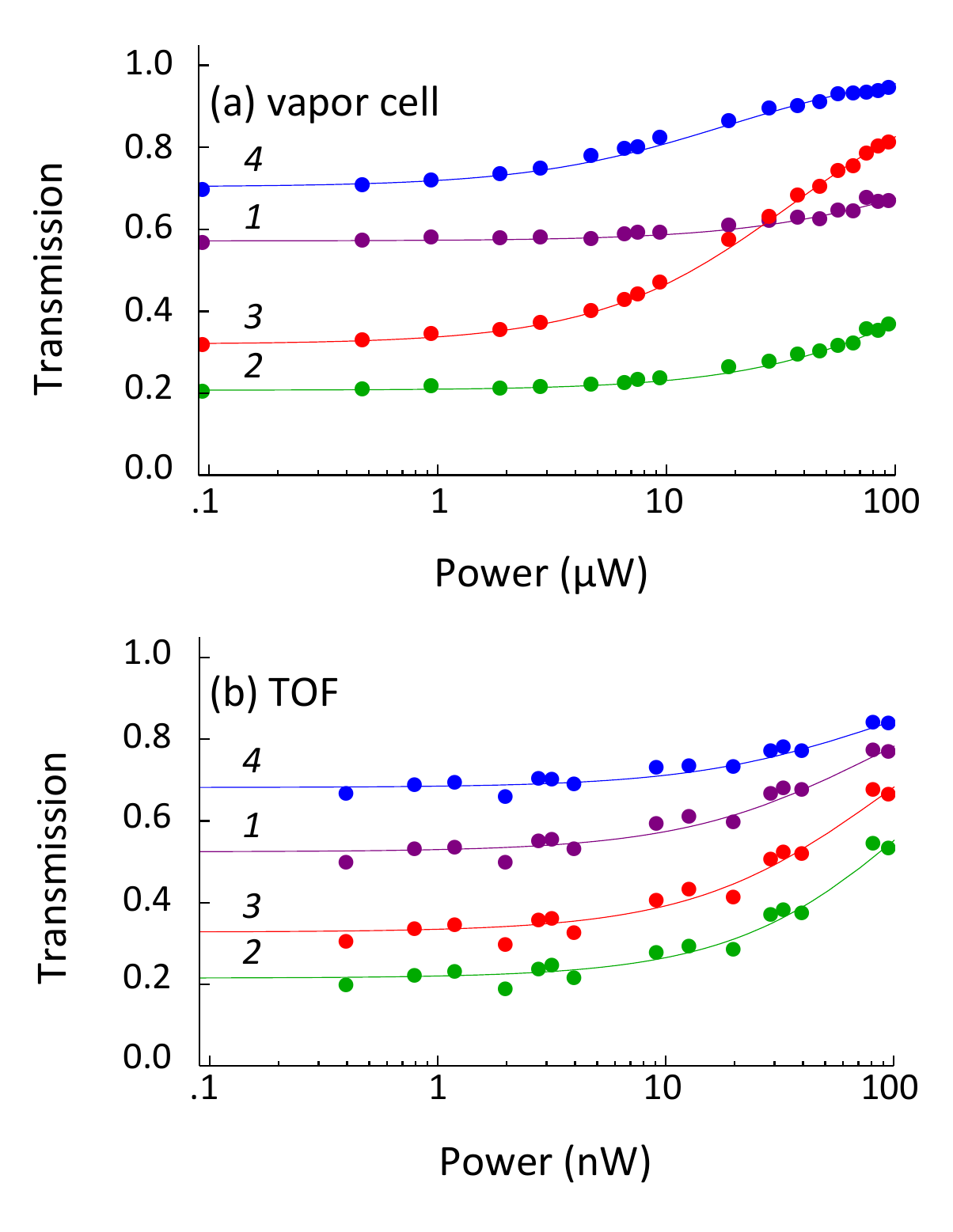}
\caption{(Color online) Comparison of the measured transmission of dips 1 - 4 at powers below saturation for (a) the vapor cell system with a large ($\sim$ 2 mm) diameter beam, and (b) the TOF system. The effects of hyperfine pumping are clearly seen in the vapor cell data of (a), where dip 3 (dip 4) appears to ``saturate'' more quickly than dip 2 (dip 1). In contrast, the effects of hyperfine pumping are not evident in the TOF system data of (b). The error bars on the data are comparable to the data point size. The solid lines are simply guides to the eye.}

\label{fig:pumping}
\end{figure}

This is clearly seen in the vapor cell data of Figure 5(a), where dip 3 (dip 4) appears to saturate much earlier than dip 2 (dip 1). 
For this particular vapor cell data set, a large $\sim$2 mm diameter beam was used to highlight the effect.  
In contrast, the TOF data in Figure 5(b) shows that all four dips begin to saturate at roughly the same input power. 
These data provide experimental confirmation that the effects of hyperfine pumping are minimized in the TOF system.


\section{Saturation modeling}
\label{sec:model}

Figure 6 shows the main results of our saturation measurements comparing the vapor cell and TOF systems. 
Although saturation has been well-studied for free-space Rb vapor cell systems \cite{siddons08,sagle96,olivares13}, analogous measurements in TOF systems have been very limited \cite{spillane08,hendrickson10,lai13}. 
By performing TOF measurements over a very wide power range, we extend the earlier results into a regime where attempting to fit the absorption data to various models becomes meaningful. 

The simplest nonlinear transmission models based on closed two-level systems typically utilize the Beer-Lambert law, $T=e^{-\alpha_{NL}*L}$, where L is the interaction length and $\alpha_{NL}$ is a power-dependent nonlinear absorption coefficient. In most cases, 
\begin{equation}
\label{eq:alpha}
\alpha_{NL} \equiv \frac{\alpha_{o}}{(1+P/P_{sat})^{r}}  ,
\end{equation} 
where P is the incident power and $\alpha_{o}$ is the linear absorption coefficient \cite{demtroderbook}. 
The exponent $r$ is equal to 1 for two-level systems dominated by homogeneous broadening mechanisms, while $r=1/2$ applies to inhomogenously (usually Doppler) broadened systems \cite{yarivbook,corneybook}.  
These models are often valid at powers much less than $P_{sat}$, but fail at higher powers due to factors such as hyperfine pumping and power broadening \cite{sherlock09}. 
Nonetheless, they can be used as a benchmark for fitting saturation data.

\begin{figure}[t]

\includegraphics[width=3.25in]{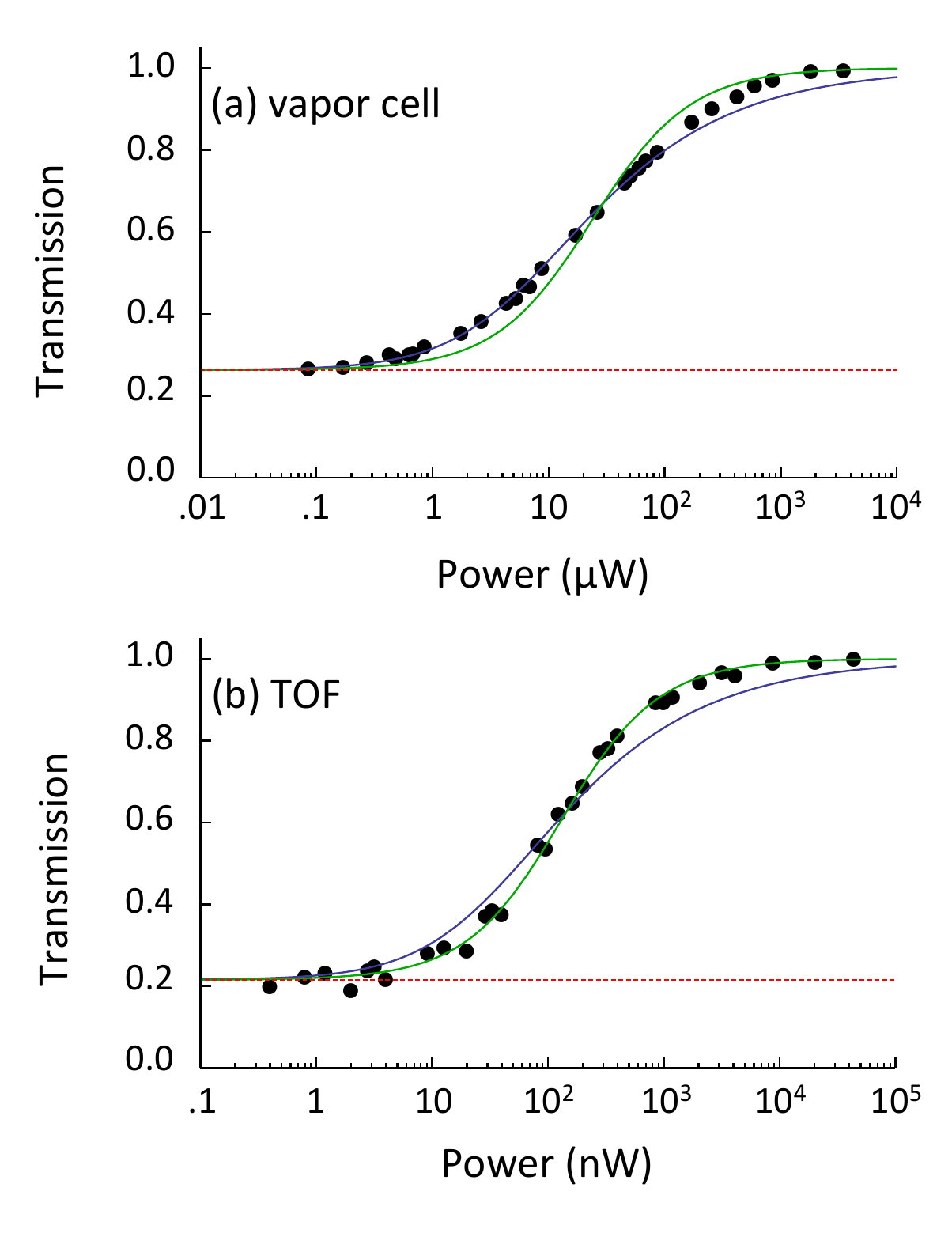}
\caption{(Color online) Comparison of the measured transmission of dip 2 for (a) the vapor cell system, and (b) the TOF system. The wide range of input powers ($\mu$W for the vapor cell, nW for the TOF) facilitate fitting the data to a simple nonlinear transmission model (described in the main text). The green curves correspond to $r=1$ and the blue curves to $r=1/2$. Note the failure of both curves in the vapor cell system in (a), and the good agreement of the green curve with the TOF data in (b). This empirical nonlinear absorption model provides a saturation power of ($61 \pm 3$) nW for the TOF system. The dashed red lines correspond to ordinary linear absorption models.}

\label{fig:saturation}
\end{figure}

Figure 6(a) illustrates this situation for the vapor cell system. 
These data show the measured transmission of dip 2 for input powers ranging from $10^{-2} \,\mu$W to $10^{4} \,\mu$W.
For this particular data set, a short focal length lens was used to focus the free-space beam to a $\sim \!20 \,\mu$m waist diameter to provide an intensity high enough to fully saturate the transition. 
The blue and green curves show fits to the data using the simple model of Eq. (\ref{eq:alpha}) with $r=1/2$ and $r=1$, respectively. 
As might be expected for the Doppler broadened vapor cell system, the blue curve  ($r=1/2$) fits the data fairly well in the low-power regime but fails at higher powers. 
The green curve ($r=1$) does not accurately fit the nonlinear transmission in either regime.

Figure 6(b) shows analogous data for the TOF system. 
Somewhat surprisingly, the simple model with $r=1$ (green curve) fits the nonlinear transmission data very well over the entire power range. 
This is consistent with the idea of increased homogeneous line broadening due to transit time effects and collisions with the TOF; however, care should be taken not to over-interpret this fit as an accurate model of the physical processes responsible for the saturation behavior. 
Rather, we simply use this model as an empirical fit to the data; in this case the model with $r=1$ gives a saturation power of $P_{sat} = (61 \pm 3)$ nW. 
The data were best fit by the same model with $r=0.88$, giving $P_{sat} = (51 \pm 5)$ nW. 
In contrast, the model with $r=1/2$ (blue curve), which gives $P_{sat} = (15 \pm 1)$ nW, does not fit the saturation behavior very well. 

We have seen the same type of fitting behavior with numerous TOF systems in varying densities of Rb \cite{hendrickson10,lai13}, and even  with TOF's in metastable Xe \cite{pittman13}. 
In all cases, the simple model with $r=1$ fits the data much better than $r=1/2$. 
Consequently, we advocate using this empirical model with values of $r \sim 1$ to provide a meaningful way to discuss and compare the saturation power for  measurements made with future TOF systems. 

It should be noted that the $P_{sat} = 61$ nW value measured in this way is comparable to a rough order of magnitude estimate of $\sim\!100$ nW obtained by using the nominal Rb saturation intensity of 1.6 mW/cm$^{2}$ \cite{steck} multiplied by the mode area seen in Figure \ref{fig:mode}, and modified by the ratio squared of the  natural line width and the measured line widths of Figure 3. 
The saturation power can also be estimated by measuring the power-broadened line width; this technique has recently been used in closely related PBGF systems \cite{slepkov10,perrella13}.

\section{Summary and Discussion}
\label{sec:summary}

TOF's suspended in atomic vapors are becoming increasingly useful for ultralow-power nonlinear optics applications \cite{spillane08,pittman13,hendrickson10}. 
Because the ability to saturate the D$_{2}$ transitions at ultralow powers is an indicator of the overall strength of the nonlinearity in these systems, understanding the saturation behavior is of great interest. 
We have experimentally studied this behavior by directly comparing power-dependent transmission in a TOF system with that seen in a standard free-space Rb vapor cell system. 
Given the difficulties in deriving an accurate theoretical model of saturated absorption \cite{stace10}, this direct experimental comparison provides two new and useful results: (1) the effects of hyperfine pumping are experimentally confirmed to be negligible in the TOF system, and (2) we have identified a simple empirical model that can be used to describe the saturation power for TOF systems in a meaningful way.  
We hope these observations provide some insight for future work on accurate modeling of the saturation behavior in this complex system, where collisions with the TOF itself and the unique optical mode shape undoubtedly play significant roles.

\section*{Acknowledgements}
We acknowledge useful conversations with G.T. Hickman and B.T. Kirby. 
This work was supported by DARPA DSO under Grant No. W31P4Q-12-1-0015.


\end{document}